\documentclass[pageno]{jpaper}

\newcommand{\asplossubmissionnumber}{XXX}

\usepackage[normalem]{ulem}
\usepackage{tikz}
\usepackage{amsmath}

\usepackage{authblk}

\usepackage{makecell}
\usepackage{multirow}
\usepackage{cite}
\usepackage{algorithm}
\usepackage{algorithmic}
\begin{document}

\title{
DeltaFS: Pursuing Zero Update Overhead via Metadata-Enabled \uline{Delta} Compression for Log-structured \uline{F}ile \uline{S}ystem on Mobile Devices}

\author[1]{Chao Wu}
\author[2]{Cheng Ji}
\author[1]{Geng Yuan}
\author[3]{Riwei Pan}
\author[4]{Weichao Guo}
\author[4]{Chao Yu}
\author[5]{Zongwei Zhu}
\author[1]{Yanzhi Wang}
\affil[1]{Department of Electrical and Computer Engineering, Northeastern University}
\affil[2]{School of Computer Science, Nanjing University of Science and Technology}
\affil[3]{Department of Computer Science, City University of Hong Kong}
\affil[4]{Guangdong Oppo Mobile Telecommunications Corp., Ltd}
\affil[5]{School of Software Engineering, University of Science and Technology of Chinese}

%\author{Chao Wu,
%Cheng Ji, 
%Geng Yuan,
%Riwei Pan,
%Weichao Guo,
%Chao Yu,
%Zongwei Zhu,
%Yanzhi Wang,
%\thanks{Chao Wu, Geng Yuan, and Yanzhi Wang are with the Department of Electrical and Computer Engineering, Northeastern University, Boston, USA. (e-mail: cha.wu@northeastern.edu, yuan.geng@northeastern.edu, yanz.wang@northeastern.edu). }
%\thanks{Weichao Guo, and Chao Yu are with Guangdong Oppo Mobile Telecommunications Corp., Ltd, Guangzhou, China. (e-mail: guoweichao@oppo.com, chao.yu@oppo.com).}
%\thanks{Cheng Ji is with School of Computer Science, Nanjing University of Science and Technology, Nanjing, China. (e-mail: cheng.ji@njust.edu.cn).}
%\thanks{Zongwei Zhu is with School of Software Engineering, University of Science and Technology of Chinese, Nanjing, China. (e-mail: zzw1988@ustc.edu.cn).}
%\thanks{Riwei Pan is with Department of Computer Science, City University of Hong Kong, Hong Kong, China. (e-mail: riweipan@cityu.edu.hk).}
%}

\date{}
\maketitle

\newcommand{\GY}[1]{\textcolor{blue}{Geng: #1}}
\newcommand{\todo}[1]{\textcolor{red}{\sf\bfseries Todo: #1}}
\newcommand{\bred}[1]{\textcolor{red}{\sf\bfseries #1}}
\newcommand{\red}[1]{\textcolor{red}{#1}}
\newcommand{\blue}[1]{\textcolor{blue}{#1}}
\newcommand{\yellow}[1]{\textcolor{yellow}{#1}}
\newcommand{\purple}[1]{\textcolor{purple}{#1}}
\newcommand{\brown}[1]{\textcolor{brown}{#1}}
\newcommand{\cross}[1]{\textcolor{red}{\sout{#1}}}

\thispagestyle{empty}

\begin{abstract}
Data compression has been widely adopted to release mobile devices from intensive write pressure. 
Delta compression is particularly promising for its high compression efficacy over conventional compression methods.
However, this method suffers from non-trivial system overheads incurred by delta maintenance and read penalty, which prevents its applicability on mobile devices.
To this end, this paper proposes DeltaFS, a metadata-enabled \uline{Delta} compression on log-structured \uline{F}ile \uline{S}ystem for mobile devices, to achieve utmost compressing efficiency and zero hardware costs.
DeltaFS smartly exploits the out-of-place updating ability of Log-structured File System (LFS) to alleviate the problems of write amplification, which is the key bottleneck for delta compression implementation. 
Specifically, DeltaFS utilizes the inline area in file inodes for delta maintenance with zero hardware cost, and integrates an inline area manage strategy to improve the utilization of constrained inline area. 
Moreover, a complimentary delta maintenance strategy is incorporated, which selectively maintains delta chunks in the main data area to break through the limitation of constrained inline area. 
Experimental results show that DeltaFS substantially reduces write traffics by up to 64.8\%, and improves the I/O performance by up to 37.3\%.\footnote{The source code will be released upon acceptance.}
\end{abstract}

\section{Introduction}
\label{sec:intro}
In the 5G era, excessive data pressure has become a huge challenge for mobile devices.
Recent report indicates that 5G users consume up to 2.7x more mobile data compared with 4G users \cite{OpenSignal}.
The intensive data pressure consumes limited hardware resources on mobile devices rapidly, which raises a huge challenge on the I/O performance and storage lifetime \cite{ji2021pattern}.
To release the mobile device from intensive write pressure, prior wisdom has made plenty of contributions.

\textbf{Limitation of prior works.}
The current F2FS file system supports the storing of file data which is smaller than 3.69KB in the inline area in file inodes \cite{lee2015f2fs, f2fsopen}.
However, the size limitation makes this technique less effective in mobile devices, e.g., only 25.5\% of files in prevalent applications in our experiments, as introduced in Tab.~\ref{tab:iopattern}, meet this requirement.
Data compression has also attracted much attention \cite{gao2019erofs, ji2021pattern}.
Some works explore conventional compressing methods, e.g., lz4, in data compression in the file system \cite{rodeh2013btrfs, Jffs2, ji2021pattern}.
However, they may suffer from several drawbacks.
First, they are less efficient in compressing small files \cite{ji2021pattern}, which dominate the mobile I/O workload \cite{7897092,lee2012smart}.
Second, the compression ratio highly depends on the data redundancy, which further depends on the data content dominated by diverse user behaviors.
This makes conventional methods result in compromised compression ratio, e.g., some can only reach a 50\% compression ratio at most \cite{lu2020adaptively}.
Third, because of the long compression/decompression time, they introduce high system overheads, including sluggish compression/decompression CPU overhead, and energy overhead.

\textbf{Potential of delta compression.}
To pursue a high compression ratio and low overhead, recent works propose to adopt the delta compression technique \cite{8110699,dsync,kim2014reducing,zhang2016reducing,zhang2019finesse,chen2020reducing}.
This technique is beneficial to yield a high compression efficacy when data are frequently updated with few content differences, as delta compression only compresses the content difference (XOR) between two data.
Prior work indicates that 70\% of updating blocks in mobile devices are updated with less than a 10\% content difference \cite{kim2014reducing}.
%A similar observation is also found in our experiments that 66.6\% database journal files are updated with less than 200Byte content differences between the original and updating data during running TWITTER application for 10 minutes.
These features of mobile I/O workload provide an opportunity to achieve a huge data reduction at a good compression ratio by using delta compression \cite{8110699}.
Moreover, because of the high compressibility, delta compression can be quickly conducted with higher compressing ratio, faster compressing speed, and lower system overheads compared with conventional compressing methods.

\textbf{Challenges.}
Albeit the outstanding compression performance, it is still challenging to implement delta compression in mobile devices.
The reasons are, the simultaneous maintenance of delta chunks (Compressed XOR) and base (original data) pages could introduce read/write amplification and extra-consumed storage space.
Prior works store delta chunks in an external buffer, e.g., NVRAM \cite{chen2018wrjfs,kim2014reducing}, or backup-battery supported DRAM \cite{8110699}.
This solution could increase the hardware cost, which is not feasible for budget mobile devices.

\textbf{Our proposal.}
Therefore, we would like to raise a question: 
\begin{quote}
\textit{Is there a way to enjoy the great benefits of delta compression without incurring any overhead? }%\textcolor{red}{put this to the end of your challenges section.}}
\end{quote}
To answer this question, we innovatively think that if we can store the data chucks by effectively exploiting the free inline area in each file node, the storage overheads could be entirely avoided. 
More importantly, the high hit rate of file inode in the main memory can help to mitigate the read/write amplification issue incurred by delta compress. 
Furthermore, we can also leverage the unique out-of-place updating policy in LFS to mitigate the write-amplification issue incurred by delta compression in conventional Journaling file systems (JFS).
Though, this seems like a promising solution, there are still several challenges need to be addressed first.
\begin{itemize}
    \item The inline area in the file system is limited, e.g., 3.69KB in F2FS, which demands a carefully designed delta-chunk management policy to reach the utmost space utilization.
%    \item Even with a dedicated delta-chunk management policy, the compressing efficacy is still bounded by the constrained inline area.
    \item Data offsets in the inline area could frequently contends with delta chunks, which incurs significant write penalty by suspending the write operation until the delta chunks are decompressed and flushed to the back-end storage.
\end{itemize}

To address these challenges, this paper proposes a novel metadata-enabled \uline{Delta} compression-based approach on log-structured \uline{F}ile \uline{S}ystem (DeltaFS) by transparently integrating the compression modules into the metadata management process.
DeltaFS
(1) integrates a delicate delta management strategy by making trade-offs between the benefits and overheads of delta compression;
(2) incorporates a machine learning-assisted delta chunk maintenance strategy, which selectively maintains delta chunks in the main data area to break through the limitation of tiny inline area.
To the best of our knowledge, DeltaFS is the first delta compression scheme with zero hardware cost in mobile devices.

We evaluated the proposed approach on a real device with F2FS, and compared it with existing approaches \cite{lee2015f2fs,ji2021pattern}.
Experimental results show that the proposed approach could substantially reduce the write volume by 55.1\% on average, meanwhile improve the average write I/O performance by 28.8\%, and read I/O performance by 25.3\%, respectively.

In summary, this paper makes the following contributions:
\begin{itemize}
    \item Comprehensively studying the features of mobile I/O workloads and log-structured file system, analyzing the advantages and drawbacks of delta compression in mobile devices.% and smartly integrates delta compression in LFS to reach the best of both worlds.
    \item Proposing a metadata-enabled delta compression technique, which leverages the wasted inline area in delta chunk maintenance to achieve aggressive compressing efficacy with marginal system overheads.
    \item Proposing a selective delta chunk management method in the data blocks to overcomes the limitation of scarce inline area and enhances the compressing efficiency.
    \item Prototyed and evaluated the proposed technique on F2FS with a real mobile device.
\end{itemize}

The remainder of this paper is as follows.
Section \ref{sec:background} reviews relevant background.
Section \ref{sec:moti} presents the motivation of this work.
Section \ref{sec:method} introduces the proposed approach.
Section \ref{sec:implement} highlights the implementing details of DeltaFS in mobile devices.
Section \ref{sec:eval} evaluates the proposed DeltaFS and analyzes the experimental results.
Section \ref{sec:related} reviews the prior related works.
Finally, Section \ref{sec:conclude} concludes this paper.

\section{Background}
\label{sec:background}
\subsection{Mobile I/O Workload}

%high write stress, small dominant, frequent update with small diff
The emerging Internet-of-Things and 5G technologies are posing mobile devices into intensive data traffic.
Excessive write I/Os contend with each other for limited hardware resources, and block read I/Os yet impact user experience \cite{nguyen2015reducing}. 
Even worse, write I/O data consumes the free space and P/E cycles of flash storage rapidly, posing the devices into higher cost for larger storage capacity.
This puts forward the necessity of releasing the write traffic for mobile devices.

One of the critical features of mobile I/O workload is that small size I/O dominates the write traffic.
Prior work indicates that small I/O requests account for at least 65\% of all write operations in mobile devices \cite{lee2012smart}.
The domination of the small I/O workloads makes conventional compression algorithms less effective.
The reason is twofold.
First, most small I/Os are randomly updated, leaving many holes in the storage space with data compression in conventional Journaing file systems.
FPC \cite{ji2021pattern} performs sequential compression for small I/Os leveraging the out-of-place updating feature of LFS, which still suffers a strict constraint of compression ratio, e.g., compressing 5 pages into 1 at most.
In contrast, delta compression, which compresses the difference between the modified and the original data of updating files, could perform a more aggressive compression, e.g., compressing 4KB data into dozen Bytes.
Second, extra mapping overhead could be incurred by traditional compressing methods, which further lowers the efficacy of these compressing methods on small I/Os.

\subsection{File System on Mobile Device}
Conventional JFS, e.g., EXT4, assign all file updating data in their last addresses, thus generating a set of random I/Os and degrading system I/O performance \cite{mathur2007new}.
To mitigate this issue, LFS propose to allocate sequential addresses to all write I/Os, and invalidate the original data addresses of file updating I/Os \cite{rosenblum1992design}.
The logical address space in LFS is divided into multiple zones.
Each zone consists of several segments.
Each segment is further divided into hundreds of blocks.
The granularity of read/write operations is block, while LFS maintains segment cleaning to reclaim invalid blocks \cite{wu2020pruning}.
\textcolor{red}{Optional: Give a small figure here or a pointer to Fig.3?}
Currently, a growing number of mobile device manufacturers are adopting LFS as their default file system for performance improvements \cite{liang2017empirical}.

\textbf{Inode Inline Area.}
To index each data offset of files, file systems (FS) maintains an inline area in each file inode structure.
A large inline area could facilitate system I/O processing, as FS could find most indexes directly from inline area instead of searching them from direct/indirect nodes.
To enjoy this benefit, modern FS supports a large inline space, e.g., 3.69KB for F2FS \cite{lee2015f2fs}.
In comparison, EXT4 only supports 256 Bytes at most.
However, a large inline area could incur space waste in small I/O dominant workload environments, which is a typical feature of mobile I/O workload \cite{lee2012smart,7897092}.
In F2FS, as an example, each 4KB block data occupies only one 4Byte data offset.
As a result, a large inline area benefits system I/O performance at the cost of wasted space in mobile devices.
To leverage the wasted inline space and fully embrace the advantages of large inline space, this paper proposes to maintain the delta chunks of delta compression in the inline area.

\subsection{Data Compression in File System}
Conventional compressing methods have been widely studied by researchers, e.g., LZO \cite{ji2021pattern}, LZ4 \cite{gao2019erofs}, LZMA \cite{lu2020adaptively}, ZLIP\cite{rodeh2013btrfs}.
However, these methods are less effective in data reduction on mobile devices \cite{ji2021pattern,lu2020adaptively}.
Recent works explore delta compression techniques to satisfy the need of high compression ratios and low system overheads \cite{8110699,dsync,zhang2016reducing,zhang2019finesse,chen2020reducing}.
Delta compression compresses the XOR value between two data.
In mobile devices, delta compression is performed by compressing the XOR between the modified page data and the original page data in each file page updating.
In this paper, \textbf{$New$ is referred to as the modified data page for short, and $Base$ is referred to as the original data page.}

One significant issue of delta compression is write amplification.
The system has to flush deltas from DRAM to flash frequently to keep data consistent.
When deltas are maintained in file inodes in JFS, FS has to flush inodes, incurring a significant write penalty.
In LFS, inodes are frequently dirtied by page writings to update the logical block address (LBA) of pages ~\cite{lee2015f2fs}.
Thanks to this mechanism, file inodes dirtied by deltas could be flushed together without introducing write penalty.
Leveraging this compression-friendly features of LFS, delta-compression could be performed aggressively with marginal system overheads, thus reaching a significantly released write pressure for mobile devices.

\section{The Feasibility and Potentiality of Delta Compression on Mobile Devices}
\label{sec:moti}

\begin{table}[!ht]
\centering
\scriptsize
\caption{I/O update (overwrite) characteristics of typical mobile application scenarios. $Vol$ is the volume of write traffic in MB,$Size$ is the percentage of files smaller than 3.69MB, $UR$ is the file updating ratio in total write traffic, $UD$ is the average updating difference between modified and original page data.}\label{tab:iopattern}
\begin{tabular}{|c|p{2.3cm}<{\centering}|c|c|c|c|}
  \hline
  App &Task&Vol&Size&UR&UD\\
     \hline
    \multirow{5}*{Gmail} &
   Launch application&0.7&100.0\% &62.4\% &31.2\% \\
   \cline{2-6}     
   &Send email with 1 image&6.3 &94.8\% &89.6\% &24.3\% \\
   \cline{2-6} 
   &Send email with 100 characters&9.0&92.4\% &91.5\% &25.3\% \\
     \cline{2-6}  
    &Receive email with 1 image&12.3 &89.1\% &71.1\% &30.2\% \\
   \cline{2-6} 
   &Receive email with 100 characters&7.2&98.6\% &62.8\% &29.0\% \\
     \hline 
     \multirow{2}*{Polish} &
    Launch application&0.1&100.0\% &78.6\% &0.9\% \\
   \cline{2-6}     
   &Edit a photo&130.6&99.9\% &51.8\% &2.7\% \\
     \hline 
    \multirow{2}*{Spotify} &
   Launch application&4.4&100.0\% &93.1\% &5.5\% \\
    \cline{2-6}
   &Listen to online music for 5 minutes&67.4&100.0\% &62.3\% &11.2\% \\
     \hline
    \multirow{8}*{Telegram} &
    Send a message with 1 image&2.4&52.9\% &85.1\% &5.2\% \\
   \cline{2-6} 
   &Send a voice message with 10 secs&3.4&88.1\% &92.7\% &3.6\% \\
    \cline{2-6} 
   &Receive a message with 100 characters&4.8&64.4\% &95.1\% &4.1\% \\
   \cline{2-6} 
   &Receive a message with 1 image&3.5&51.7\% &84.4\% &1.6\% \\
   \cline{2-6} 
   &Receive a voice message with 10 secs&4.7&72.1\% &86.9\% &2.5\% \\
     \hline 
    \multirow{4}*{Twitter} &
  Launch application&9.2&100.0\%  &61.4\% &25.2\% \\
   \cline{2-6} 
   &Post a comment with 100 characters&12.5&100.0\% &92.9\% &11.1\% \\
      \cline{2-6}
   &Post 1 image&15.8&100.0\% &84.5\% &8.3\% \\
      \cline{2-6}
   &Watch news for 5 minutes&117.5&99.2\% &75.6\% &14.4\% \\
     \hline 
   \multirow{8}*{Wechat} &
    Launch application&2.0&97.0\%  &67.1\% &14.4\% \\
   \cline{2-6} 
   &Send a message with 100 characters&2.3&94.6\% &75.9\% &21.7\% \\
    \cline{2-6}
   &Receive a message with 100 characters&1.8&97.7\% &76.5\% &22.3\% \\
    \cline{2-6}
   &Receive a voice message with 10 secs&2.8&85.4\% &55.3\% &9.2\% \\
    \cline{2-6}
   &Post 1 image&0.7&92.1\% &61.2\% &14.7\% \\
     \hline    
   Zoom & Attend a could meeting for 5 minutes&207.4&82.0\% &92.5\% &13.1\% \\
     \hline 
\end{tabular}
\end{table}

To fully exploit the superiority of delta compression on mobile devices, there are several preferences for the file of compressed data.
First, the file size has to be in a specific range.
For example, the size needs to be smaller than 3.69MB in a 3.69KB inline area.
For files larger than 3.69MB, the inline area in file inodes is filled with data offsets, without spare space to accommodate deltas.
Second, the file has to be frequently updated, to maximize the compressing efficiency.
Third, the content difference between the modified and original data of the updating file should be small.

%Delta-compression is fundamentally suitable for frequent updating files with tiny updating difference.
\textbf{Investigation of Delta Compression Possibility.}
To study the benefit of delta-compression on mobile devices, we first evaluated 7 popular applications with 13 common user behaviors.
The experimental results are shown in Table.~\ref{tab:iopattern}.
Most files in the selected tasks, 89.7\% on average, are smaller than 3.69MB.
File updating pages take account of 77.1\% over the total write traffic on average, with only 13.8\% average updating differences.
These results indicate the high potential of delta-compression in write traffic release.
Among involved application launching scenarios, SQLite files take account of 88.1\% in all files, which meets the observation of prior works \cite{tuan2016io, 8110699}.
The second target scenario is typing characters, as the system needs to update the SQLite files to save the characters.
SQLite files take account for 41.9\% of all updating files.
The third target scenario is multimedia file-involved user behaviors, e.g., send/receive/edit images, or voices.
%A set of SQLite files have to be updated to serve these user behaviors, which takes account for 46.2\% in all files.
Meanwhile, temporary files takes account for 15.5\% in all files in multimedia involved scenarios, which are updated frequently to record the contents of browsing multimedia files from the Internet.

\begin{figure}[!ht]
        \centering
        \includegraphics[width=3in]{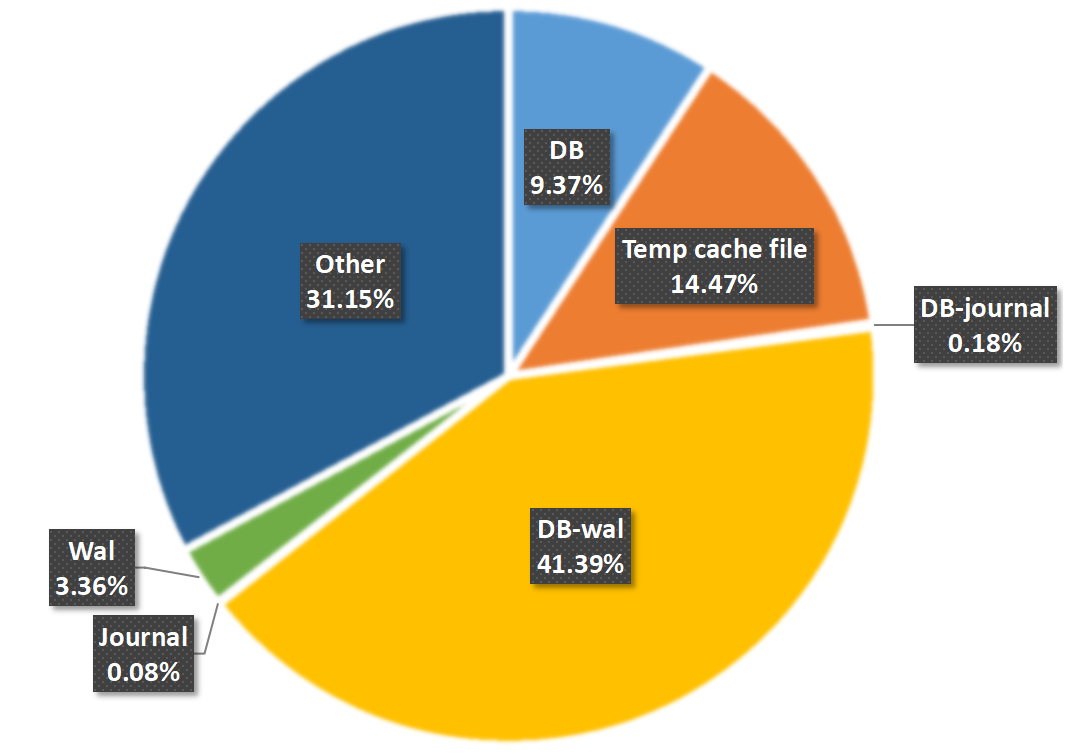}
        \caption{Composition of the inspected mobile files.} 
        \label{fig:composition}
\end{figure}

To further study the target files of delta-compression, we collected the number of all involved file types in all scenarios in Table.~\ref{tab:iopattern}.
As shown in Figure.~\ref{fig:composition}, there are two major file types which meet the requirements of delta-compression.
The first type is the SQLite database files, with the suffix of db, db-journal, db-wal, wal, and SQLite.
In total, the SQLite files take a portion of 53.6\% in all files.
This result meets the observations of prior works \cite{lee2001differential,8110699}.
The second type is temporary files in cache, mostly starting with todelete\_ or random characters or temp/tmp in the suffix, which takes a portion of 13.5\% in all files.
%These temporary files will be removed soon by the system or users.
%However, the creation and storage of these files could sluggish the system I/O performance and consume the precious flash lifetime resources.
%Hence, the compression on these files could benefit the device efficiently.

\textbf{Opportunity for Merging Delta Chunks with Metadata.} 
\begin{figure}[!ht]
        \centering
        \includegraphics[width=3in]{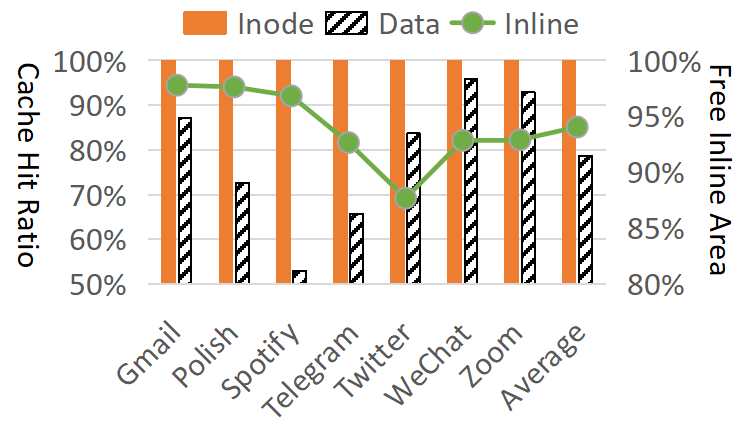}
        \caption{Opportunity for delta-inlining.} 
        \label{fig:inlinewaste}
\end{figure}
Although beneficial to mobile devices, the significant overheads incurred by delta compression hinders its widespread adoption.
By separating compressed data with delta and base, delta compression could incur read/write amplification and extra-consumed storage space.
%It is promising to mitigate these issues by accommodating delta chunks in the free inline area in file inodes.
%Log-structured file systems have a relatively large inline area, which can be taken advantage of for storing additional data.
Fig.~\ref{fig:inlinewaste} represents the opportunity of merging delta chunks in file inodes.
The vertical left axis represents the cache hit rate of file inodes and data, to show the advantage of merging delta chunks with file inodes.
The hit rates of file inodes accessed by involved applications approach to 100\% (99.89\% at lowest, 99.97\% on average).
In contrast, the hit rates of file data is much lower (52.9\% at lowest, 78.7\% on average).
%The reason is straightforward, since the system needs to access file inode to track all data in the file.
Since the operating system always needs file inode before any file data tracking, it is promising to mitigate compression overheads by accommodating delta chunks in the free inline area in file inodes.

Delta-inling could eliminate the extra-consumed storage issue of delta compression, since most inline area in file inodes on mobile devices are free because of the small-dominant mobile workloads.
The vertical axis in Fig.~\ref{fig:inlinewaste} represents the average free inline area of file inodes.
The experimental results are evaluated by the average free inline space over 3.69KB, which is the size of the inline area in each file inode.
For large files whose inline area is filled with data offsets, the free space is 0 in calculation.
94.0\% inline area of involved files are free in all evaluated applications.
Intuitively, the inline area is highly utilized for large file accessing, while the free inline space has a high potential to be leveraged for small files.
Combining these delta chucks with metadata updates can help to reduce write frequency.

In summary, small updates dominate the overall write traffic, and delta compression has a high potential to reduce such write stress.
Conventional delta-compression techniques do not consider the features of mobile file systems, which ignore the opportunity of exploiting the file system mechanism for the best performance enhancement.
For mobile devices, it strongly motivates to introduce an efficient delta and base data maintenance approach, aiming at maximizing the
compression benefit with a marginal overhead.

\section{Methodology}
This section presents the design principles and details of our proposed DeltaFS.

\label{sec:method}
\begin{figure}[!ht]
        \centering
        \includegraphics[width=3.3in]{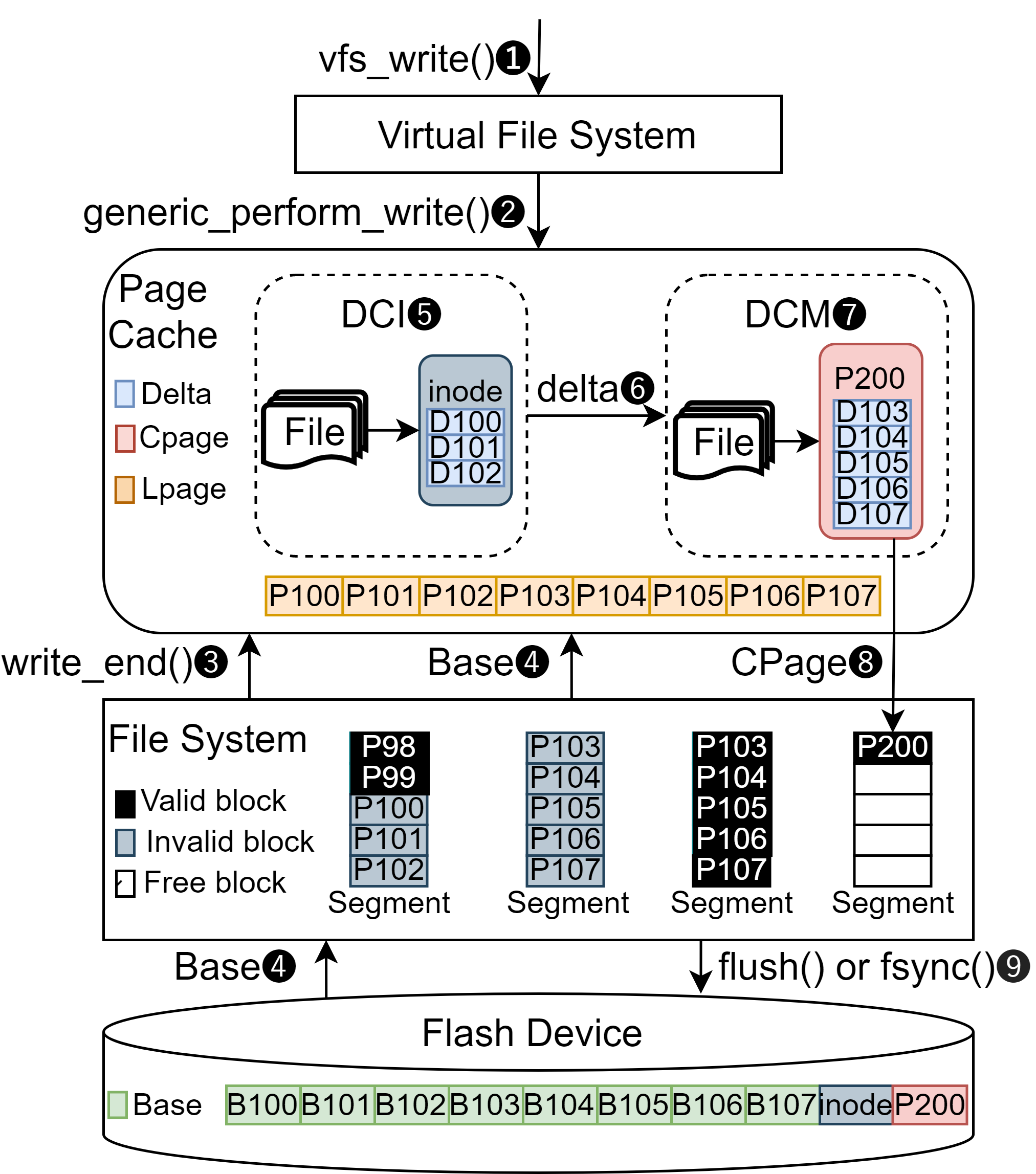}
        \caption{Architecture of DeltaFS. 8 data pages (Lpage, P100-107) in a file are updated and buffered in the page cache \textcircled{1}\textcircled{2}\textcircled{3}. DCI reads the base \textcircled{4} to compress the data, and maintains delta chunks (D100-102) in the inline area \textcircled{5}. When the inline area is exhausted, DCM compacts remaining delta chunks (D103-107) into one block, buffers the block as a page (Cpage, P200) in the page cache \textcircled{6}\textcircled{7}\textcircled{8}. Finally, P200 and the inode is synchronized into the storage device \textcircled{9}. } 
        \label{fig:arch}
\end{figure}
%This section presents the architecture and the details of the proposed compression approach. 

\subsection{Delta Compression Inlining (DCI)}
\begin{figure}[!ht]
        \centering
        \includegraphics[width=3.3in]{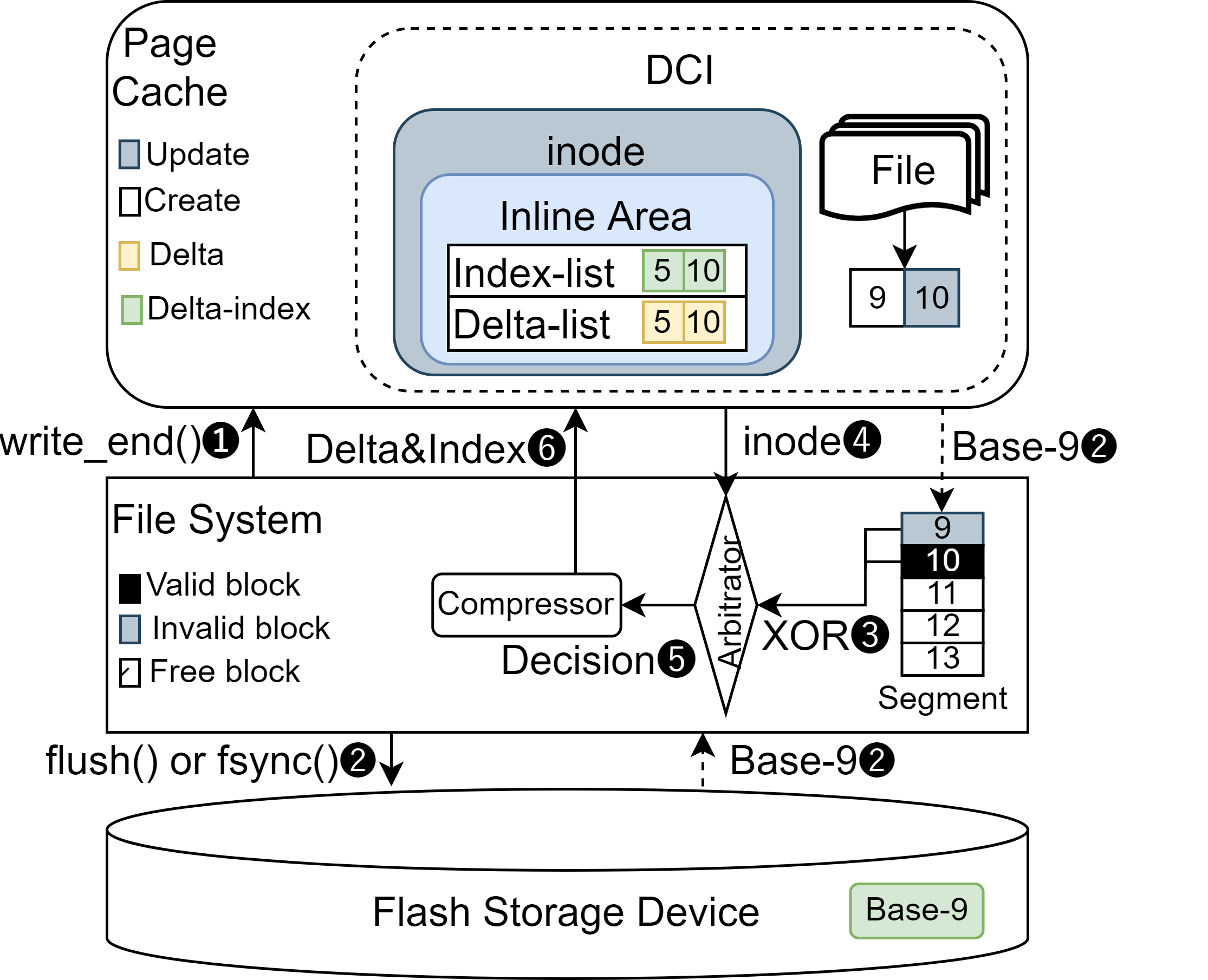}
        \caption{Workflow of DCI.  \texttt{Update} and \texttt{Create} represents the existing data page updating and the appended data page allocation of a file, respectively.} 
        \label{fig:dciarch}
\end{figure}
DeltaFS mainly incorporates delta chunks in the file node (referred to as \textbf{DCI}, \uline{D}elta \uline{C}hunk \uline{I}nlining), achieving aggressive data compression via zero-cost software manner.

\subsubsection{Workflow of DCI:}
Figure.~\ref{fig:dciarch} shows the workflow of DCI.
In DCI, delta compression is performed in the unit of one block to keep align with the existing file system.
For example, F2FS maintains the logical address space with 4KB data blocks as the minimal granularity.
Once file updating is performed, $New$ (page 10) is buffered in the page cache \textcircled{1}.
DCI reads $Base$ from page cache \textcircled{2} (dashed line) or flash device \textcircled{2} (solid line).
Then, arbitrator in DCI compares $Base$ and $New$ to calculate the compression ratio of the XOR between both \textcircled{3}.
DCI reads file inode to perceive the information of the inline area \textcircled{4}.
According to the size of the compressed XOR and the status of the inline area, the arbitrator decides to perform delta compression or not \textcircled{5}.
For uncompressed data, page 10 is marked dirty and pending for system flushing.
Otherwise, the compressor compresses the XOR, pushes the delta (compressed XOR) into the inline area together with the index and size of the it, dirties the inode and sends it back to the page cache \textcircled{6}.
Note that, \texttt{DCI maintains a compressing tag in file inode to identify each compressed file}.
In file readings, DeltaFS first wanders the page cache to read $New$.
If page miss happens for compressed pages, DCI simply reads $Base$ from the flash and delta chunks from file inode, and decompresses these pages.
Another notable thing is, \texttt{DeltaFS does not update $Base$ in each data compression.}
When the delta size is larger than the threshold (\textless{256} Bytes in DCI as introduced later), DeltaFS simply writes the uncompressed data, deletes the corresponding delta chunk in the inline area, and invalidates $Base$ for space reclaiming.

The process for compressed file data in DeltaFS finishes when the delta is maintained in the inline area in the corresponding dirtied file inode, which eliminates the necessity of flushing $New$ into the flash storage.
In comparison, the current LFS has to proceed the page writing by marking $New$ as dirty page, and waiting for the next system flushing to synchronize the dirty page into storage.
By this aggressive manner, DeltaFS could reduce the write traffic significantly, thus reaching an improved system I/O performance, and prolonged flash storage lifetime.
\subsubsection{Metadata Inline Area Exploitation.}
%Which data to be compressed and inlined? Set thresholds for hot, warm and cold data, benchmark different thresholds for each type.
%Which delta to be evicted?
%Maybe can configure LRU, compress rate order list and ARC, and set it configurable?
%Each 4KB file compress.
%If delta larger than threshold, not compress and inline.
%Else, check if inline area has the delta file.
%Inline area maintained with LRU.
%If does, invalidate and evict the old delta.
%If inline area has no enough free space, evict.
%If inline area does not have delta, insert, check if has enough area, if not evict.

\begin{figure}[!ht]
        \centering
        \includegraphics[width=3in]{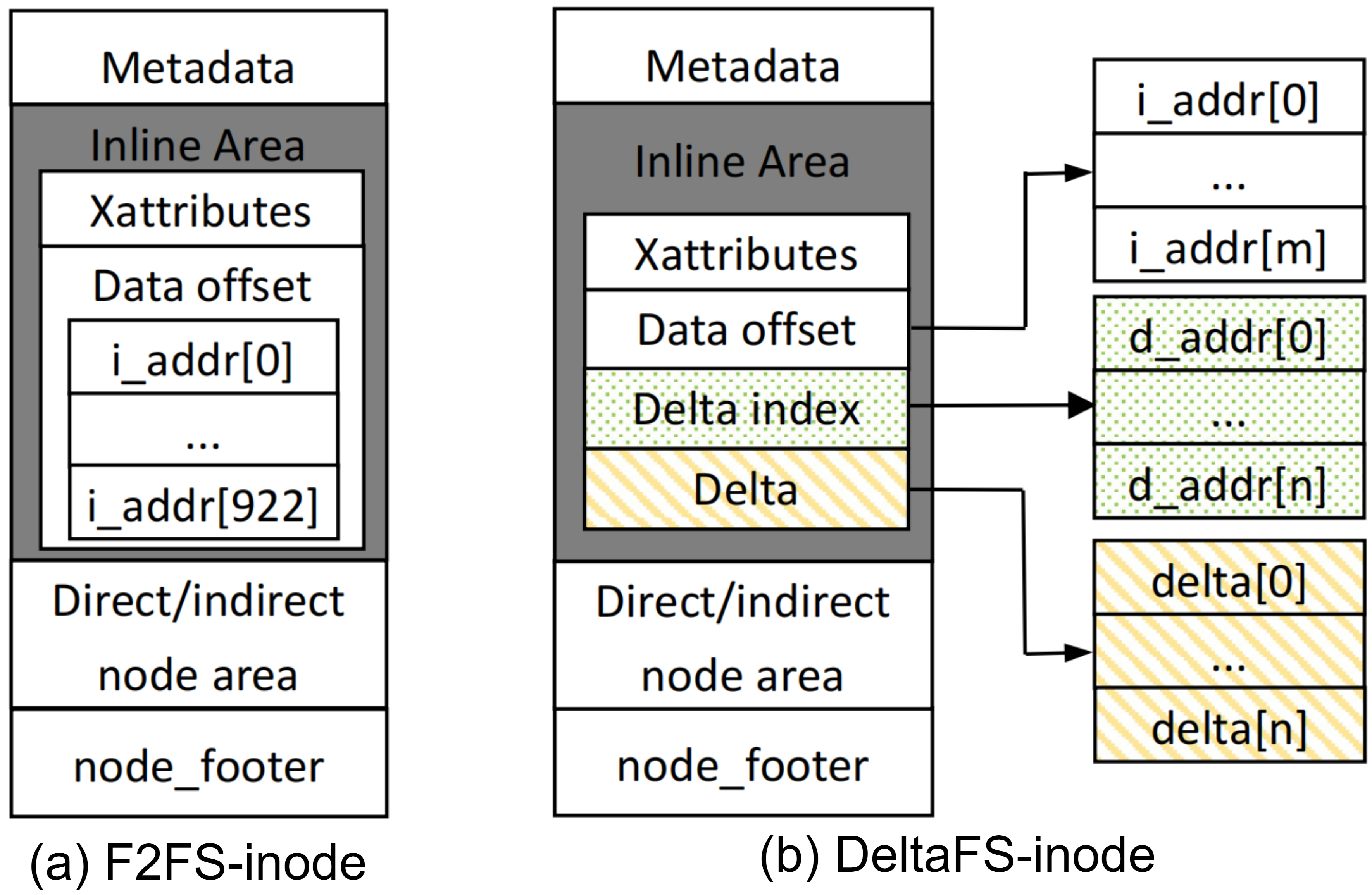}
        \caption{Comparison of inode layout between original FS and the proposed DeltaFS.} 
        \label{fig:inline}
\end{figure}
\begin{figure}[!ht]
        \centering
        \includegraphics[width=2.3in]{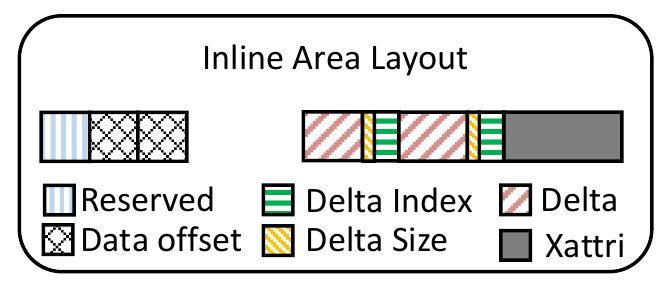}
        \caption{Inline area layout. Besides the Xattr in the tail of the inline area, the index of the first delta chunk is inserted (takes 2Byte); the size of the delta chunk is placed next (takes 1Byte to record delta chunks smaller than 256Byte); the delta chunk is inserted at the end from right to left.} 
        \label{fig:inlinemanage}
\end{figure}

% Under extreme space constraint， e.g., less than 4KB-> without inducing extra costs.
DeltaFS exploits the existing inline area of LFS to store the compressed delta chunks. While conceptually simple, the implementation is quite challenging, since the available space resources in the inline area is still constrained, e.g., less than 4KB.
Under this extreme space constraint, it is challenging to reach high space utilization without inducing extra costs.
Fig.~\ref{fig:inline} compares the inline area layout between F2FS (an example of LFS) and DeltaFS.
In F2FS, 923 data offsets are maintained, which takes 3.69KB in inline area (4Byte per data offset) in a 4KB inode.
Together with the direct and indirect nodes, F2FS could track a 3.93TB file.
Moreover, F2FS supports the maintenance of file \textit{xattributes} in the inline area, which stores 50 data offsets.
In this case, 3.488KB free space is left to maintain the data offsets.
Since small I/O dominates the mobile I/O workload \cite{lee2012smart,7897092}, inline area with large free space could incur a significant space waste.

To enable delta compression and utilize a large inline area, DeltaFS maintains the delta chunks in the inline area, which complicates the management of the inline area.
As shown in Fig.~\ref{fig:inline}, besides the \textit{xattributes} and the data offsets, the inline area in DeltaFS maintains the delta indexes and the delta chunks.
The number of delta chunks maintained in the inline area depends on both file size and the size of each delta chunk.
For example, 256 data offsets are required for a 1MB file, which takes 1KB free space in the inline area.
Suppose the average compression ratio for all compressible data pages in this file is 98.9\% (the delta size is 44Byte, which is the minimum size for delta compression), 56 deltas could be maintained in the inline area.

Fig.~\ref{fig:inlinemanage} shows the inline area layout in DeltaFS.
Two lists are maintained in the inline area, delta-list and delta-index-list.
Each delta index maps to the data offset of the corresponding delta.
Two available formats could be adopted to maintain the mapping between the delta and the corresponding block data.
In the first format, all deltas are aligned in the unit of one data offset (4Byte), while the padding space will be zero-filled.
The belonging delta index is appended adjacent to the delta, taking 2Byte in the inline area.
In this case, one delta chunk could occupy 6Byte in the inline area.
However, this method is only suitable for compressed codes with a size of 3-4Byte.
Otherwise, space waste is incurred for redundant delta index maintenance.
In the second format, deltas are aligned in their true value in Bytes, while the delta-index-list maintains both the data offset and the size of the corresponding delta.
With this format, each delta with the minimum size takes the space of 47Byte in the inline area, while \texttt{52 deltas could be maintained for the example discussed above.}
As shown in Fig.~\ref{fig:inlinemanage}, the current data offsets are maintained in the head of the inline area together with one reserved offset.
The file \textit{xattributes} which takes 50 offsets is maintained in the tail of the inline area.
In DeltaFS, the delta together with its index and size are inserted in the tail of the inline area, adjacent to the file \textit{Xattributes} following the order from tail to head.
Each time when delta index meets the data offset, the inline area replacement policy is triggered.

\textbf{Internal Node Data Replacement.}
The retrieval of compressed data in DeltaFS incurs significant system overhead.
Data decompression needs to read the base either from the page cache or from the back-end storage.
Besides, the decompression process consumes computing resources.
Finally, the decompressed data should be marked as dirty in the page cache, pending for the next system flush to write back to the flash storage, which incurs write amplification.

In contrast, delta replacement in the inline area spares more inline area to accommodate the subsequent delta.
Literally, the benefit gained by each delta replacement depends on the spared size as well as the saved I/O latency between compressed and uncompressed data.
Moreover, the compressed data accommodated in the spared inline area could reduce the consumed flash P/E cycles.

The trade-off between system overhead and benefit incurred by delta replacement should be carefully balanced.
Assume the average size of compressed XOR together with its index is $\alpha$, the average write I/O latency of compressed and uncompressed write I/O is $\beta$ and $\gamma$ respectively.
The benefit gained by each delta replacement could be calculated with Formation.~\ref{benefit}.
\begin{equation}\label{benefit}
  Ben=\frac{RI}{\alpha} \times (\gamma-\beta)
\end{equation}
In which $RI$ is the spared size in the inline area.
Note that, the term of reduced P/E cycles is not introduced in the benefit formulation for calculating convenience, meanwhile guaranteeing the benefit gained by each delta replacement is larger than the system overhead it incurs.

The system overhead of delta replacement could be calculated with Formulation.~\ref{overhead}.
\begin{equation}\label{overhead}
  OH=\gamma + \varepsilon + HR \times \lambda + (1-HR) \times \delta
\end{equation}
In which $OH$ is the system overhead incurred by delta replacement, $\varepsilon$ is the decompression latency, $HR$ is the hit rate of base in the page cache, $\lambda$ is the read latency of hit base page in the page cache and $\delta$ is the latency of miss base page.
For example, the compressed delta size $\alpha$ is 72Byte, the average I/O latency of compressed and uncompressed write I/O $\beta$ and $\gamma$ are 25.3us and 954us, respectively.
The decompression latency $\varepsilon$ is 6.9us, base page hit rate $HR$ is 74.1\%, the read latency of hit and miss base page $\lambda$ and $\delta$ are 9.2us and 250us respectively.
All these data are evaluated on a real device whose configuration is described in Sec.~\ref{sec:eval}.
In this case, the spared inline area for the delta replacement should be larger than 81Byte to guarantee the benefit gained is larger than the overhead incurred.
Hence, the size of the delta which is inserting into the inline area should be smaller by 77Byte at least than the delta which is about to be evicted.

\textbf{When to compress.} Theoretically, delta compression could be performed either when the data is accessing the page cache or when the data is flushing from the page cache to the back-end flash storage.
For the latter solution, excessive memory resource could be consumed.
In each file updating, LFS invalidates the original data page, which will be reclaimed by the segment cleaning soon.
Once performing delta compression in data flushing, DeltaFS needs to keep the original data page in the page cache as the base, which could pollute the precious page cache resource.
Hence, DeltaFS performs delta compression in page cache accessing, in function \textit{write\_end()} in the file system layer.

DCI is advanced by leveraging file inode in delta compression, to reach high compressing effect with mitigated read/write amplification and space waste.
However, the write traffic release of DCI is still bounded by the constrained inline area.
Meanwhile, delta chunks evicted from DCI could degrade the QoS (quality-of-service) performance of mobile devices, as the system needs to decompress and flush them to guarantee data consistency.
To address these issues, a complementary delta chunk maintenance scheme, DCM (\uline{D}elta \uline{C}hunk \uline{M}aintenance for the main data area), is introduced.
%\begin{figure}[!ht]
%        \centering
%        \includegraphics[width=3in]{asplos21-templates/pic/contend.drawio.png}
%        \caption{An example of the contention between data offset and delta chunks.}
%        \label{fig:contend}
%\end{figure}
%Figure.~\ref{fig:contend} describes an example of the contention.
%In the initial stage, four delta chunks are maintained in the inline area, while the system tries to write the file data with the offset 800 in the file.
%As the location of the data offset is occupied by delta 2 in the inline area, DeltaFS has to evict delta 2-4 to spare the occupied space for data 800.

\subsection{Delta-Chunk Maintenance (DCM)}
%compress target
\begin{figure}[!ht]
        \centering
        \includegraphics[width=3.3in]{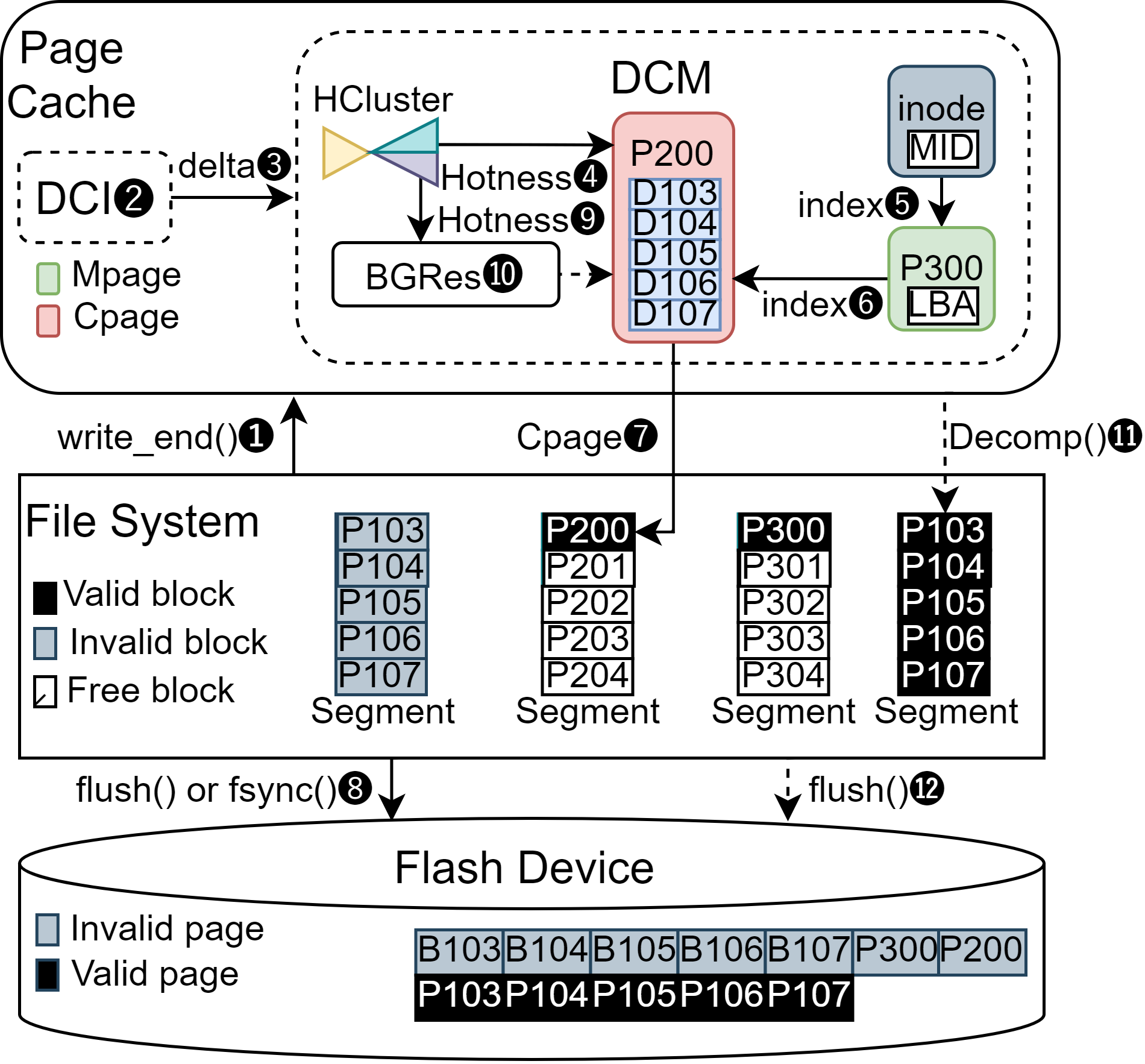}
        \caption{Workflow of DCM. Mpage is meta-page which records the logical addresses of compressed pages, Cpage is compact page which contains compressed delta chunks.} 
        \label{fig:dcmarch}
\end{figure}
\subsubsection{Workflow of DCM.}\label{subsec:drawbackdcm}
Fig.~\ref{fig:dcmarch} presents the working process of DCM.
DCM handles delta chunks generated from DCI \textcircled{1}\textcircled{2}\textcircled{3}, which means $New$ of these delta chunks should be clean pages in the page cache, while $Base$ should be in the flash.
When the inline area in file inode is exhausted, DCM first performs file hotness clustering (HCluster) for the corresponding file \textcircled{4}.
For write-hot-read-cold files, DCM allocates a logical block (P200) for delta chunks \textcircled{7}, updates the logical block address (LBA) in the meta-page (P300, introduced in Sec.~\ref{sec:implement}) \textcircled{6}, whose LBA is recorded in file inode (MID) \textcircled{5}.
Finally, system flush or fsync writes the compact page (P200) \textcircled{7}, meta-page (P300) and file inode to the back-end flash device.
In system idle period, e,g,.background segment cleaning \cite{wu2020pruning}, the background data restoration mechanism (BGRes) is triggered, which first performs Hcluster to detect the hotness status of the file \textcircled{9}, and decides to decompress delta chunks of the file or not \textcircled{10}.
If yes, new logical blocks will be allocated for all decompressed data \textcircled{11}, which will be marked as dirty pages and flushed to the flash device later \textcircled{12}.

\textbf{Data types for compression.}
DCM maintains two types of delta chunks in main data area.
First, the evicted delta chunks in internal node data replacement strategy, and the delta chunks contending with the data offsets in the inline area.
Second, file updating data that could have been compressed while the inline area is exhausted.
For the former one, delta chunks are inserted into selected pages directly.
For the latter one, the compressing data should be carefully selected, to mitigate the system overhead incurred by main segment delta maintenance.

It is non-trivial to maintain delta chunks in the main data area.
Literally, delta compression separates each compressed page data with base and delta chunk.
In data restoration, the system needs to read both the base and delta chunk, which could incur read amplification.
DCI addresses this issue by merging the delta chunk with file inode, whose hit rate is extremely high in the memory, which is yet challenging for DCM.
Hence, DCM only targets on write-hot-read-cold data in data compression.
For the same reason, DCM could incur write amplification.
To balance the trade-off between benefits and overheads in data compression, DCM targets on data which could be compressed at least from 5 pages to 1 page, which is the highest compression ratio in \cite{ji2021pattern}.
Finally, DCM only compresses clean data to strictly keep the latest version of data in the page cache and the base in the flash device, as DCI does.

\textbf{Learning-based file hotness clustering (Hcluster).}
Multiple works have been proposed to identify the hotness of data.
Most works perform data clustering in page granularity, according to the temporal/spatial locality of read/write I/O requests \cite{8942140, 10.1145/3319647.3325840, 10.1145/3319647.3325829}.
This solution might be sub-optimal if implementing in the file system, as it is the file accessing latency which decides the user experience on the device, which is in turn dominated by the longest page accessing latency of all belonging data pages.
One file might contain data pages with different hotness, while the accessing speed of the file could be decided by that of cold data. 
Some works propose to cluster page data based on semantic/program context \cite{227784, lee2015f2fs, 6808214, 216874} or file types \cite{210522}, while files in the same semantic/program or with the same type might have different hotness.
Hence, \texttt{DeltaFS perform data clustering in file granularity}, referring to \cite{zhang2021automatic}.
%analyzed in detail, foreground should cluster with enlarged distance to get larger hot data, type? 

%\begin{figure}[!ht]
%        \centering
%        \includegraphics[width=2in]{asplos21-templates/pic/hcluster.drawio.png}
%        \caption{File characters as the input of Hcluster.} 
%        \label{fig:hcluster}
%\end{figure}
Hcluster (machine learning-assisted \uline{H}otness \uline{cluster}ing takes the read/write count of each file as the input.
Intuitively, the input could be accumulated and recorded in each file inode.
However, this cannot represents the recent hotness of each file.
According to the definition of \texttt{temporal locality}, the recent accessed file have high potential to be accessed soon.
Hence, Hcluster records the read/write count of each file within a time window $T$, which is a hyper-parameter.
%$App$ is the name of the current application, identified by the dentry path in file inode.
%$App$ is labeled accumulatively, while background applications are labeled as 0.
%$File Type$ includes SQlite files, temporary files, multimedia files, cookies, text files, apk files and other files.
%The file types are identified by the suffix of the file stored in file inode in the kernel, labeled from 0 to 6.
%$freq\_r$ and $freq\_w$ are the read/write frequency of the file, referred to as the temporal locality.
%$size$ is the number of compressed pages in the file, which is input to make sure the compressed page number in the file is more than 5.
We adopt kmeans++ as the clustering algorithm, exporting the file hotness (read-hot-write-cold, read-cold-write-hot, read-cold-write-cold, or read-hot-write-hot) as the output \cite{ilprints778}.
The output will be taken into consideration in delta maintenance decision making, as shown in step 4 and 9 in Fig.~\ref{fig:dcmarch}.
Note that, Hcluster is an online learning scheme.
After each clustering, Hcluster updates the centroids of each cluster, to adapt to the time-varying file characters.
%inode chain in sb or cp2
%hotness change
%size change
%
%mapping

\textbf{Background data restoration (BGRes).}
As introduced above, DCM could introduce system overheads.
Although DeltaFS addresses this issue by carefully selected compressing target data, this issue could still be significant as file size/hotness could vary with user behaviors.
DCM integrates BGRes (\uline{B}ack\uline{G}round data \uline{Res}toration) to adapt to the time-varying file characters.
BGRes maintains an dual-linked inode list, whose head is recorded in the super block.
Each inode in the list corresponds to the file whose deltas are maintained by DCM.
In the extra-attr area of each inode, an inode number is kept to track to the next inode in the list.

In system idle time, e.g., background segment cleaning period, BGRes traverses files in the list.
For each file, HCluster is first performed, to identify the hotness of the file.
If the file is read-hot, BGRes decompresses all delta chunks of the file that maintained by DCM.
Otherwise, BGRes detects the delta chunk number in DCM, and decompresses all delta chunks if the number is less than 5.
The decompressed data pages will be marked as dirty in the page cache, pending for the next system flushing to write them back to the flash device.
In this manner, DCM guarantees a long-term aggressive benefit of delta compression.

\subsection{Read/write process.}

\begin{algorithm} 
	\caption{Write process} 
	\label{algo:inlinewrite} 
	\begin{algorithmic}
		\REQUIRE $page,\ inode$ 
		\ENSURE $compress\ completion\ status$ 
		\IF{$page\_dirty(page)$}
		\STATE $return\ FAIL$
	    \ENDIF
		\STATE $base=read\_base(inode, pgae->index)$
		\STATE $delta=compress(page,base)$
		\IF{$inline\_free\_space(inode)>sizeof(delta)$}
		\STATE $insert\_delta\_inline(inode, delta)$
		\STATE $set\_inode\_flag(inode)$
		\ELSE
		\IF{$inline\_node\_replace(inode, delta)==FAIL$} 
		\IF{$file\_hotness\_cluster(inode)$}
		\STATE $alloc\_data\_block\_for\_delta(inode,delta)$
		\ELSE
		\STATE $set\_page\_dirty(page)$
		\STATE $return\ FAIL$
		\ENDIF
		\ENDIF
		\ENDIF
		\STATE $return\ SUCCESS$ 
	\end{algorithmic} 
\end{algorithm}
		
\begin{algorithm} 
	\caption{Read process} 
	\label{algo:inlineread} 
	\begin{algorithmic}
		\REQUIRE $inode,\ index$ 
		\ENSURE $data\ for\ reading$ 
		\IF{$compress\_type(inode) == INLINE$} 
		\STATE $delta=traverse\_inline\_for\_delta(inode, index)$
		\ELSIF{$compress\_type(inode) == MAIN$}
		\STATE $delta=read\_delta\_in\_block(inode, index)$
		\ENDIF
		\STATE $base=read\_base(inode, data\_offset)$ 
		\STATE $XOR=decompress(delta)$
		\STATE $data=calc\_xor(base,XOR)$
		\STATE $return\ data$ 	
	\end{algorithmic} 
\end{algorithm}
Algorithm.~\ref{algo:inlinewrite} and ~\ref{algo:inlineread} describe the write/read process of DeltaFS.
One trick in DCI writing is, when a compressed page is being updated, DeltaFS will not refer to the flash device to read the base.
Instead, it reads the corresponding delta from file inode, decompresses delta and calculates the XOR between delta and the latest page data before updating to get the base.
In function $inline\_node\_replace()$, DCM is triggered to handle the evicted delta chunks, or the current delta chunk if no replacement happens.
Another notable thing is, if the offset in the inline area is occupied by delta chunks in page offset updating, DeltaFS evicts contending delta chunks and trigger DCM to accommodate these delta chunks in the main data area.
This process happens in page syncing to the flash, which is not described in Algorithm.~\ref{algo:inlinewrite} owning to the page limit.

%conclusion
\section{Implementation Remarks}\label{sec:implement}

\textcolor{black}{The proposed approach can be implemented as a compatible module to existing mobile log-structured file systems.}

\textbf{Data alignment.}
DeltaFS strictly maintains two data versions for each compressed data page.
$New$ pages are marked as clean in the page cache to be prevented from flushing to the storage.
Meanwhile, $Base$ pages are maintained in the flash storage.
The objective is, $New$ pages in the page cache could benefit read performance, which is crucial to the user experience.
$Base$ pages in the flash could guarantee data consistency.
To achieve this objective, \texttt{DeltaFS only compresses clean pages in the page cache}.

\textbf{D2D (delta to data) mapping in DCI.}
DCI maintains the mapping between delta and data in the inline area, as shown in Fig.~\ref{fig:inlinemanage}.
The logical address of each compressed page is maintained in the corresponding data offset in the inline area, which indexes to both $Base$ in the storage and $New$ in the page cache.
To get a delta chunk for specific compressed data, DeltaFS could traverse the delta indexes in the inline area, to find the delta whose index equals to the offset of the target data.
As the constrained inline area could not accommodate many delta chunks, the traversing overhead should be marginal.

\begin{figure}[!ht]
        \centering
        \includegraphics[width=3.3in]{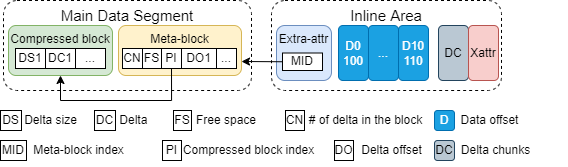}
        \caption{D2D (delta chunks to data) Mapping in DCM.} 
        \label{fig:maindelta}
\end{figure}
\textbf{D2D (delta to data) mapping in DCM.}
DCM separates each compressed data into delta chunk and base, which needs to be indexed simultaneously.
Same with DCI, DCM maintains the base offset in the corresponding data offset in the inline area.
To index the delta chunks, DCM configures the extra-attr area in the inline space, which locates in the beginning of the inline area, as shown in Fig.~\ref{fig:maindelta}.
One viable solution for D2D mapping is to keep the alignment of delta chunks in each data block in the main data area the same as in the inline area in DCI.
However, DCM might maintains multiple compressed blocks in the main data area.
For this solution, DCM might have to traverse multiple delta chunks in all compressed blocks to find a specific delta chunk, which could incur significant system overhead.
To mitigate this issue, DCM proposes to take one data block for each file as the meta-block, which maintains the mapping pair between each index of the compressed page (PI, each takes 4 Bytes) and the offsets of all delta chunks in that block (DO, each takes 2 Bytes).
Ahead of each mapping pair, 1 Byte space is reserved to record the number of all delta chunks in the compressed block (CN).
Next, 2 Bytes are used to record the available free space in the compressed block (FS).
The logical address of the meta-block (MID) is recorded in the extra-attr area, taking 4 Bytes in the inline area.
In this manner, delta chunks in the compressed block are aligned in the order of: delta chunk size (2 Bytes), delta from head to tail.
Note that, the threshold of all maintained delta chunks in DCM differs from that in DCI, which is \textless{256} Bytes.
As explained earlier, there are plenty of space in main data area, thus DCM targets on data which could be compressed from 5 pages to 1 page to achieve more aggressive compression efficacy than conventional compression \cite{ji2021pattern}.
Hence, 2 Bytes are reserved in the compressed blocks for each delta chunk to record its size.

\subsection{Data consistency}
Delta-compression might pose the system into data corruption, since the delta and base of each compressed block data are maintained separately.
If a system crash happens when the delta and base of one block data in the flash storage are inconsistent, the OS fails to recover this data.
The current LFS, e.g., F2FS, flushes dirty inodes and data separately, which cannot guarantee the consistency between deltas and bases.
In DeltaFS, we strictly keeps the base pages in the flash, while the current flushing mechanism in LFS helps to synchronize file inode to the flash.
In this manner, file data could be restored by decompressing all compressed data with delta and base.
Moreover, the meta-pages and compact pages are write back to the flash immediately once they are updated.
This method ensures data consistency with marginal overheads.

\section{Evaluation}
\label{sec:eval}
\subsection{Experiment Setup}
\subsubsection{Experimental Environment.}
The experimental environment is based on Oneplus 8T smartphone with 8-core Qualcomm Snapdragon 865 CPU, 8GB DRAM memory, and 128GB UFS flash storage.
The AOSP of the device is LineageOS \cite{lineage}.
The version of Android and Linux kernel are 11 and 4.19, respectively.
LZO \cite{lzo} is adopted as the default compressing algorithm in DeltaFS.
F2FS is configured as the default file system.
As the current F2FS switches to IPU (in-place updating) mode when space utilization is high or \textit{fdatasync} is triggered for small files, we configure F2FS with out-of-place updating only.
The reason is DeltaFS aims to implement delta compression in LFS, while IPU is adopted specifically in F2FS.

\subsubsection{I/O Workload Environment.}
\begin{table}[!htbp]
\centering
\caption{I/O workload characteristics. $Vol$ is the volume of data writing to the storage disk in GB, $IO$ is the number of I/O requests, $RW$ is the ratio of write I/O over total I/O, $CCR$ is the compression ratio of conventional compressing methods, and $DCR$ is the compressing ratio of delta compression.}\label{tab:workload}
\begin{tabular}{|c|c|c|c|c|c|}
\hline
App &  Vol & IO & RW & CCR & DCR   \\
\hline
   GM &  3.12& 60091 &96.1\% & 37.7\%& 96.9\%\\
\hline
  PS & 0.39&11391 & 94.8\%& 53.1\%&97.3\% \\
\hline
 ST & 0.27& 33035 & 98.7\%& 28.2\%&97.6\%\\
\hline
  TG &0.45 &12400  &65.6\% & 24.6\%& 98.4\%\\
\hline
  TW  & 0.88 & 19231 &61.8\%& 30.0\%&97.1\%\\
 \hline
WC &1.38 & 32805 &75.5\% &39.4\% &98.2\% \\
\hline
ZM & 0.38& 2041 & 72.7\%& 44.7\%& 96.5\%\\
\hline
\end{tabular}
\end{table}
DeltaFS is evaluated with a set of popular applications, including gmail (\textbf{GM}), polish (\textbf{PS}), spotify (\textbf{ST}), telegram (\textbf{TG}), twitter (\textbf{TW}), wechat (\textbf{WC}), and zoom (\textbf{ZM}).
These applications are performed with a set of common user scenarios, including sending/receiving message (\textbf{TG,WC}), listening to web music (\textbf{ST}), sending/receiving email (\textbf{GM}), editing photos (\textbf{PS}), watching web news ( (\textbf{TW})), and attending cloud meetings (\textbf{ZM}).
Each application is cold-launched and played for 30 minutes, with a mixture of user behaviors as shown in Table.~\ref{tab:iopattern}.
An accurate trace replay tool is developed to track the file characteristics.
The tracked file characteristics include file name, file inode number, file operation, operation time, file data content and the offset of each data.
Next, the collected file operations are replayed with a user-level process, strictly following the order of each file data operation.
Table.~\ref{tab:workload} describes the I/O characteristics of our collected traces, where $Vol$ is the total writing volume in GB, $I/O$ is the number of I/O requests, $RW$ is the ratio of write I/O volume over the total I/O, $CCR$ is the average compressing ratio of conventional compressing methods, while $DCR$ is the average compressing ratio of delta compression.
As shown, most involved user behaviors of selected applications are write-intensive, for which delta compression exhibits a superior compressing efficacy over conventional compressing methods.

\subsubsection{SOTA Works.}
We compare DeltaFS with multiple works, including the original F2FS without data compression (\textbf{F2FS-NODC}), and the current F2FS with data compression (\textbf{F2FS-DC}).
Moreover, we deem \textbf{FPC} \cite{ji2021pattern} as the baseline, which is a SOTA work that proposes to compress random I/O requests with conventional compression technique.
Note that, SOTA delta-compression-oriented works \cite{8110699,kim2014reducing} are not evaluated.
These works adopt an external buffer cache in delta accommodation, which is not practical for mobile devices for cost, size and I/O management complexity reasons.

We compare SOTA works with sole \textbf{DCI}, \textbf{DCM}, and \textbf{DeltaFS}, respectively.
In DCI, we decompress all evicted delta chunks and flush the data pages immediately when delta replacement or contention happens between delta chunks and data offset.
In DCM, we compress the selected data pages and maintain the delta chunks in the main data area directly, as discussed in Sec.~\ref{sec:method}.
In application launching time evaluation, each application is cold-launched for 5 times.
We take the average launching time as the final results.

\subsection{Experimental Results}
\subsubsection{Write Stress Release.}
\begin{figure}[!ht]
        \centering
        \includegraphics[width=3in]{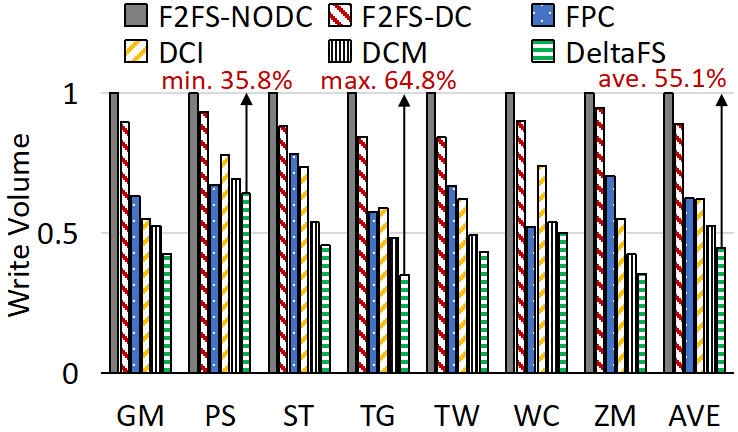}
        \caption{Normalized write volume of various approaches.} 
        \label{fig:wr}
\end{figure}

Fig.~\ref{fig:wr} shows the write volume of various approaches, normalized to the write volume of F2FS-NODC as shown in Table.~\ref{tab:workload}.
The lower value in the figure is the better, which means more write traffic is reduced.
On average, DeltaFS can reduce the write volume by 55.1\%.
For TG application, DeltaFS could release the write stress by up to 64.8\%.
Both DCI and DCM play a critical role in write stress release.
As shown, DCI can reduce the write volume by 37.9\% on average, which shows a similar compressing efficacy over FPC.
%For some applications, e.g., WC, DCI shows a similar compressing efficacy over FPC (29.2\% lower for WC).
The constrained inline area resource limits the delta-inlining technique in DCI to achieve an aggressive compressing efficacy.
In contrast, DCM could reach a higher write volume reduction effect, by 47.4\% on average, by holding delta chunks in the main data area.
However, the result is achieved at the cost of reduced page cache performance, which will be introduced later.
DeltaFS takes the advantage of both DCI and DCM, by maintaining most delta chunks in the inline area, and selectively maintains few delta chunks in the main data area.
As a result, DeltaFS could achieve enhanced compressing efficacy.

In contrast, traditional compressing methods show a sub-optimal compressing effect.
F2FS-DC reduces the write volume by 10.9\% on average, which could only compresses sequential pages in each file.
FPC mitigates this issue by compressing random pages, thus reducing write traffic by 37.6\% on average over F2FS-NODC.
However, the compressing efficacy of FPC is still bounded by the relatively low compressing ratio of conventional compressing methods.
Moreover, DeltaFS is partially orthogonal with conventional compressing methods, e.g., FPC.
Developers could choose to compress the appending file data with conventional compressing methods, meanwhile compressing updating file data with DeltaFS, to reach an even more aggressive compressing efficacy.
However, this is beyond the scope of this work.

In summary, DeltaFS could significantly release mobile devices from intensive write stress, by holding most delta chunks in the inline area, and selectively maintaining read-cold-write-hot delta chunks in the main data area when the inline resource is exhausted.
The released write traffic helps to prolong the lifespan of flash device, and improve the I/O performance, as indicated by prior works \cite{ji2021pattern, zhang2019finesse}.

\subsubsection{I/O Performance.}

\begin{figure}[!ht]
        \centering
        \includegraphics[width=3in]{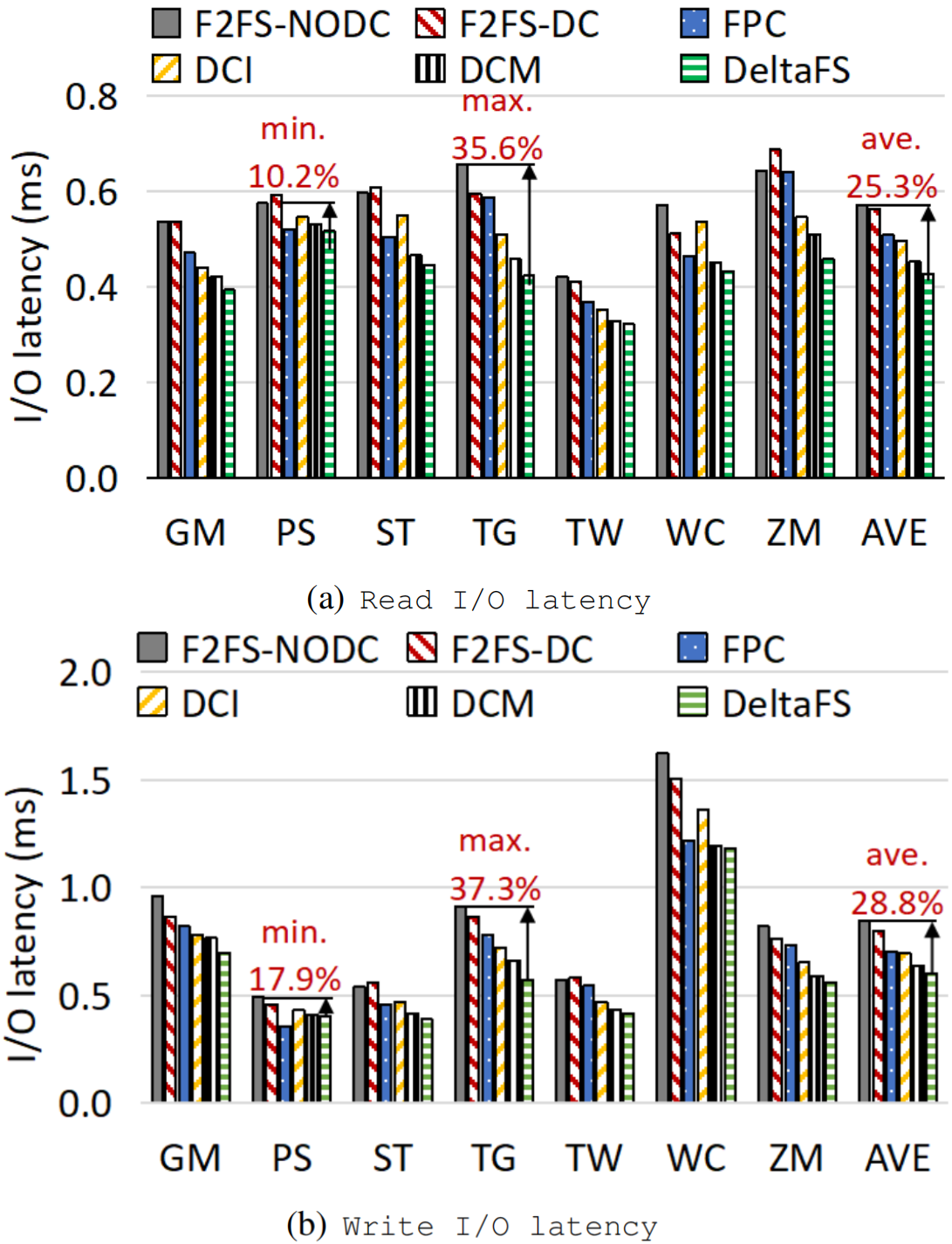}
        \caption{I/O latency of various approaches.}
        \label{fig:iolate}
\end{figure}
The released write stress of mobile devices could directly improve the I/O performance owning to the reduced I/O contention, which has been pointed out by prior works \cite{ji2021pattern, chen2016ordermergededup, 277842}.
Fig.~\ref{fig:iolate} describes the I/O latency of various approaches.
DeltaFS shows an optimal I/O performance, 25.3\% and 28.8\% faster than F2FS-NODC on average for reading and writing I/Os, respectively.
The reason is the best compressing efficacy of DeltaFS, which leads to the optimal write traffic reduction.
One typical example is the I/O latency of TG.
For TG, the average latency of reading I/O is reduced by DeltaFS by 35.6\%, and that of writing I/O is reduced by 37.3\%.
The significantly reduced write traffic of TG (64.8\% reduced) could mitigate the resource conflict between I/O requests, thus contributing to the I/O performance.
%The improved I/O performance could benefit the app launching performance, which has been indicated by prior work \cite{9211463}.

DCM could improve the I/O performance significantly, owning to the reduced write traffic.
The average latency of reading I/O is reduced by DCM by 20.8\%, and that of writing I/O is reduced by 24.5\%.
However, DCM cannot mitigate the low cache hit rate issue incurred by delta compression (59.9\% average cache hit rate), as described in Fig.~\ref{fig:chr}.
In contrast, DCI shows a slightly lower I/O performance.
Compared with F2FS-NODC, DCI reduces the average latency of reading I/O by 13.0\%, and that of writing I/O by 17.5\%.
However, DCI could deliver an excellent page cache performance (78.0\% average cache hit rate), as shown in Fig.~\ref{fig:chr}.
The improved cache hit rate by DCI could benefit the system performance, owning to the significantly faster cache accessing speed over flash devices.

In summary, DeltaFS manages to improve the I/O performance by reducing the write traffic on mobile devices, which in turn leads to the reduction of I/O contention.
The improved I/O performance could benefit the user perceived performance, which has been indicated by prior work \cite{9211463}.

\subsubsection{Cache Hit Rate.}
\begin{figure}[!ht]
        \centering
        \includegraphics[width=3in]{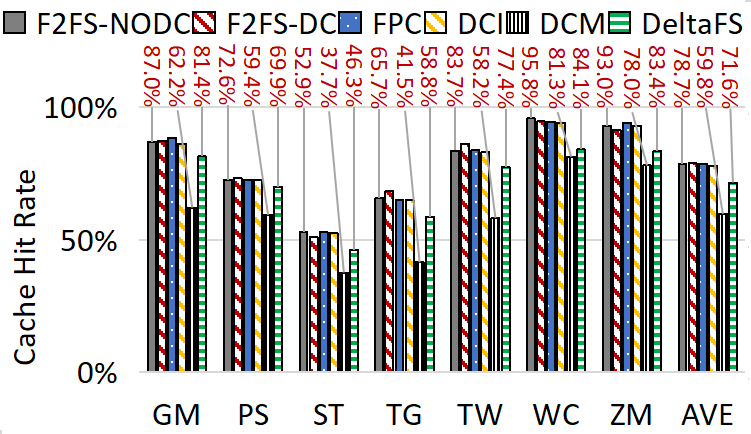}
        \caption{Cache hit rate of page data.} 
        \label{fig:chr}
\end{figure}
To figure out the issue of read/write amplification in delta compression, we evaluated the hit rate of file data in the page cache.
For DCI, DCM, and DeltaFS, both the cache hit rate of data pages and delta chunks are evaluated.
As shown in Fig.~\ref{fig:chr}, DeltaFS shows a good page cache performance (71.6\% hit rate on average).
As analyzed above, DCI maintains delta chunks in file inodes, which keep residing in the memory during file accessing.
For this reason, the average cache hit rate of DCI is similar with F2FS-NODC (78.0\% over 78.7\%).
Same with DCI, DeltaFS advocates the delta-inlining technique in the maintenance of most delta chunks, thus leading to a close cache hit rate with F2FS-NODC.
Take GM as an example, DCI accommodates 44.9\% delta chunks in file inodes.
As DeltaFS reduces the write traffic of GM by 57.3\%, only 12.3\% delta chunks are maintained in the main data area in DeltaFS.
This explains the close cache hit rate of GM between DCI (86.2\%) and DeltaFS (81.4\%).

In contrast, DCM shows the lowest cache hit rate, 59.8\% on average.
Even with selective data compression, the maintenance of delta chunks in the main area in DCM could still hurt the cache hit rate.
DeltaFS addresses this issue by hosting most delta chunks in file inode, as described above.
As shown in Fig.~\ref{fig:wr}, DCI reduces the average write volume by 37.9\%, while the value of DeltaFS is 55.1\%.
Which means in DeltaFS, only 17.2\% delta chunks area maintained in the main data area.
Hence, DeltaFS is capable of mitigating the read/write amplification issue of DCM to a large extend.

\subsubsection{Evaluation on Hcluster and BGRes.}
\begin{figure*}[!ht]
        \centering
        \includegraphics[width=6.6in, height=1.6in]{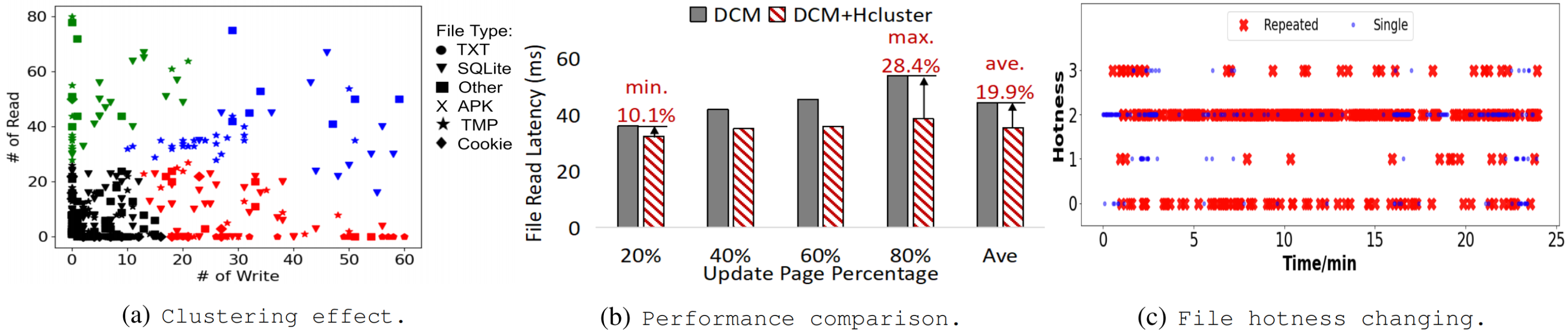}
        \caption{Micro-benchmark on the effect of Hcluster and BGRes in DCM.} 
        \label{fig:microbench}
\end{figure*}
To mitigate the system overhead incurred by main data area delta-maintenance, DeltaFS introduces Hcluster and BERes in DCM.
Fig.~\ref{fig:microbench} (a) depicts the clustering effect of files within a period in our experiments, as a snapshot.
We observed that, the hotness of SQLite files disperse among all clusters, which indicates the inefficiency of file type-based clustering.
Moreover, most temporary files, which are one of main compressing targets in DeltaFS as discussed in Sec.~\ref{sec:moti}, are read hot.
The maintenance of read-hot delta chunks in main data area might lead to read performance penalty, which is not acceptable for mobile devices since read performance dominates the user experience.
This poses the necessity of Hclustering adoption.

To quantify the impact of Hcluster on read performance, a set of experiments are evaluated.
In the experiments, the mobile device is configured as DCM with and without Hcluster, respectively.
To reflect the common effect of the delta compression for mobile devices,a small file usage scenario is simulated and evaluated: an 1MB new file is created and then updated its multiple pages with a few content difference (between 1 - 100 bytes) to make these pages compressed.
The number of updated pages is set to be 20\%, 40\%, 60\%, and 80\%, respectively.
Then, random read is performed on the file with \texttt{Mobibench} for 1000 times \cite{jeong2013androstep}, to make the file being clustered by Hcluster as read-hot.
The average file read latency are reported.
Fig.~\ref{fig:microbench} (b) shows the results.
Hcluster does help to improve the file read latency.
With the increasing number of compressed pages in the file, the disparity of read performance between DCM with and without Hcluster increases apparently. 
The reason is straight-forward.
With the increasing number of compressed pages, more hot delta chunks are maintained in the main data area by DCM if without Hcluster.
This could exacerbate the read/write amplification issue of delta compression, which in turn leading to degraded read performance.

Fig.~\ref{fig:microbench} (c) describes the file hotness changing w.r.t. time in minutes, which are the same snapshot as in Fig.~\ref{fig:microbench} (a).
Hotness 1,2,3,4 denote read-cold-write-cold, read-cold-write-hot, read-hot-write-cold, and read-hot-write-hot, respectively.
\texttt{Repeat} means these files are repetitively accessed in multiple minutes with hotness changed, while \texttt{Single} means the opposite.
In this experiments, the hotness of 56.4\% files among all change over time.
For files whose hotness change from read-cold to read-hot, the compressed delta chunks of them could incur read penalty, as discussed in Fig.~\ref{fig:microbench} (b).
This poses the necessity of BGRes adoption.

\subsection{Sensitivity Study}
DeltaFS could be sensitive to the size threshold of delta chunks in both DCI and DCM.
However, the size threshold of delta chunks in DCI, which is 256Byte, is actually most viable.
To enlarge the threshold, DeltaFS has to spend 1 extra Byte for each delta chunk, which violates the cost-benefit for precious inline area resource.
Hence, we evaluated the sensitivity of DeltaFS on the size threshold of delta chunks in DCM, which denotes the number of delta chunks in each file when DeltaFS decides to maintain these delta chunks in the main data area or not.
In the experiments above, this threshold is configured as 5 to compete the maximum compressing efficacy of conventional compressing method (FPC).

We take TG as the example to study the sensitivity, as DeltaFS delivers an excellent compressing efficacy for TG.
\begin{figure}[!ht]
        \centering
        \includegraphics[width=3in]{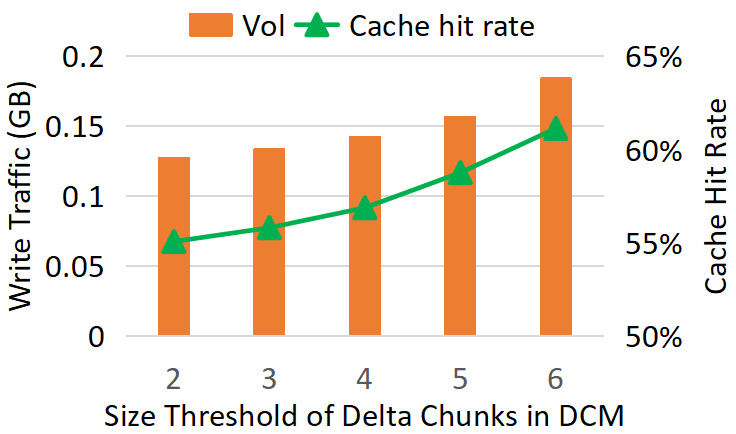}
        \caption{Sensitivity study on the size threshold of delta chunks in DCM.}
        \label{fig:sensitive}
\end{figure}
Fig.~\ref{fig:sensitive} shows the results, from which we can get several observations.
First, there is a trade-off between write traffic reduction and cache hit rate incurred by the size threshold.
A larger threshold leads to higher cache hit rate and lower write traffic reduction effect, and vice versa.
This is because, a smaller size threshold make DeltaFS maintains more delta chunks in the main data area, which could exacerbate the read/write amplification issue.
We configure the threshold as 5 to get a Pareto-optimal write traffic reduction and cache hit rate, which is not always the best choice.
For example, system designers could choose to enlarge the threshold for some critical file accessing, or only maintains deltas evicted/replaced from DCI in the main data area, to reach a better performance.
Second, a small size threshold cannot guarantee a significantly improved compressing efficacy.
For example, the write traffic is only reduced by 1.5\% for threshold 2 over threshold 3.
In contrast, the write traffic is reduced by 6.1\% for threshold 5 over threshold 6.
The reason is, DCM maintains a meta-page to record the mapping information of each file.
For files with only two delta chunks, DCM cannot benefit the system when size threshold is 2.

\subsection{Overhead Analysis} \label{subsec:overhead}
\subsubsection{Compressing/Decompressing Overhead.}
One main advantage of delta compression is the marginal compressing/decompressing overhead over conventional compressing methods.
To quantize the advantage, we evaluated the average compressing/decompressing latency of various approaches.
In our experiments, the average compressing/decompressing latency of conventional compressing method is 48.8 and 45.2 microseconds, respectively.
In contrast, the average compressing/decompressing latency of delta compression in DeltaFS in 7.4 and 8.3 microseconds.
As the average read I/O latency in FPC is 507.5 microseconds, the decompressing latency takes a portion of 8.9\% over the read I/O latency.
With the increasing number of compressed pages, the impact of compressing/decompressing latency on system performance could be obvious.

\subsubsection{Overhead of DCI.}
%DCI delta traverse
DCI could significantly degrade the QoS (quality-of-service) performance of mobile devices, which is because of the contention between delta chunks and data offsets, and the internal node data replacement policy.
The evicted/replaced delta chunks need to be decompressed and flushed immediately to guarantee data consistency, during which the system has to suspend the page writing operation whose data offset is about to be updated in the inline area.
As a result, the contention/replacement leads to an extremely long page writing latency.

\begin{figure}[!ht]
        \centering
        \includegraphics[width=3in]{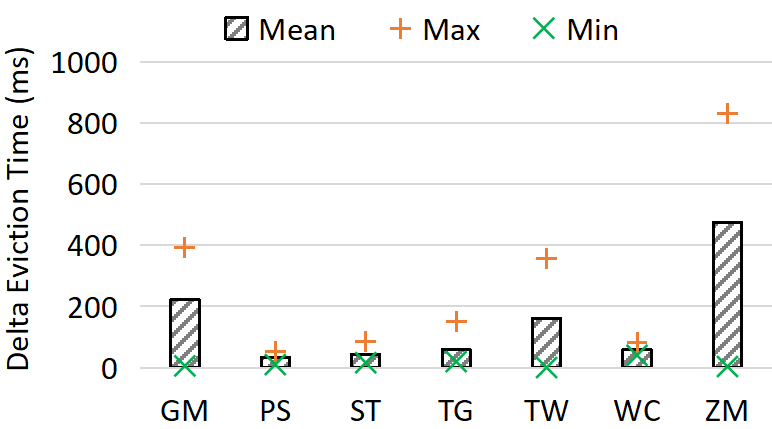}
        \caption{Delta eviction time of each app.}
        \label{fig:contend}
\end{figure}
As data offset updating and delta replacement happen before the write page is packed into I/O requests, we cannot evaluate the worst-case I/O latency to study the impact of both on the QoS performance.
Instead, we evaluated the processing time it takes for DCI to decompress and flush all evicted/contending delta chunks.
Fig.~\ref{fig:contend} shows the results.
The Average processing time ranges from 42-475ms, while the longest latency is up to 830ms.
In our experiments, this extremely long processing time of delta restoration could lead to frame-blocking/dropping, or sometime I/O task hanging for watchdog barking.
This is unacceptable for user experience-oriented mobile devices.
To mitigate this issue meanwhile breaking through the limitation of scarce inline area, DeltaFS incorporates the DCM scheme, which allocates data blocks in main data area for delta maintenance.

DCI might have to traverse multiple delta chunks in file inodes to find the target delta chunk.
The traversing overhead takes 0.2 microseconds on average, and 2.4 microseconds at maximum, which is marginal over the average I/O latency.

\subsubsection{Overhead of DCM.}
DCM could incur several system overheads.
The first overhead is incurred by meta-pages and compact pages.
For each compressed file, DCM takes 1 page as the meta-page the record the mapping information.
As DCM compresses at least 5 pages into 1 page, the flushing of meta-pages and compact pages should not incur read/write amplification.
The only overhead it takes  is the storage space in flash, which takes 7.1MB for all files from involved apps in our experiments, which is marginal over 4.2GB files generated from all the apps in our experiments.
The second overhead is the clustering latency in HCluster, which takes 1.2 microseconds on average, and 2.0 microseconds at maximum.
This is marginal over the average I/O latency.
The third overhead is incurred by BGRes.
As shown in Fig.~\ref{fig:contend}, DeltaFS amortizes the overhead of delta eviction/replacement in the background.
As the background processes in LFS is interruptable, e.g., background segment cleaning, BGRes could not interfere the system performance.

In summary, DeltaFS manages to reduce the write traffic of mobile devices with marginal system overheads.
.

\section{Related Work}
\label{sec:related}
In this section, we focus on the related works on conventional data compression and delta compression.
\subsection{Conventional Data Compression}
Ji {\it et al.} \cite{ji2021pattern} propose FPC, which performs foreground sequential compression on write-intensive barely-read I/O data leveraging the out-of-place updating feature of LFS, and background compression to re-organize read critical blocks of executable files.
FPC achieved to reduce the write traffic of SQLite to 80\% at most, e.g., compressing 5 logical blocks to 1 physical block, while this paper offered a much larger write reduction (boosting the compression ratio to as high as 97.6\%), which overcomes the limitation of conventional compression in terms of compression ratio.
Gao {\it et al.} \cite{gao2019erofs} propose EROFS, a new compression-friendly read-only file system that leverages fixed-sized output compression and memory-efficient decompression to achieve high performance on mobile devices.
Zhang {\it et al.} \cite{zhangxuebin} propose to accelerate the executable files compression with hardware in the file system in mobile devices.
These works cannot overcome the constraint of conventional compressing methods, yet fail to perform aggressive data compression on mobile devices.
\subsection{Delta Compression}
To mitigate the detrimental effect of conventional compression, several works adopt the delta compression technique \cite{kim2014reducing, 8110699}.
In the storage side, Kang {\it et al.} \cite{kang2013x} propose X-FTL to adopt log buffer as a journaling space to reduce SQLite journaling.
Park  {\it et al.} \cite{277826} propose a learning-based delta compression in post-deduplication in SSDs.
Chang {\it et al.} \cite{zhang2016reducing} propose a delta compression-based approach for all updated files in SLC area of SSDs for prolonging the lifespan of the SLC area.
Wu {\it et al.} \cite{wu2012delta} propose delta-FTL to compress the updated journaling files with delta compression and store the delta in SSDs.
These works are orthogonal to our DeltaFS, as we focus on the data compression in the file system layer on mobile devices.
In the host side, Kang {\it et al.} \cite{lee2001differential} adopt differential logging to reduce the write stress of in-memory DB systems.
Other works \cite{kim2014reducing,8110699,oh2015sqlite} propose external non-volatile buffers as the log buffer, which could increase the cost of mobile devices, thus less attractive to mobile devices.

While there exist plenty of compression works, this study explores the deep exploitation of the delta compression technique for mobile devices by tightly and transparently integrating the compression modules into the file system metadata mechanism.
First, DeltaFS maintains most delta chunks in the inline area without the maintenance cost of external buffers, e.g., NVRAM, capacitor-backed DRAM/SRAM.
Second,  DeltaFS eliminates the read amplification by fetching most delta chunks from the memory-resident file inodes.
Third, our DeltaFS advocates both advantages of LFS and delta compression to mitigate the issue of write amplification, while these works are not efficient in this issue.

\section{Conclusion}
\label{sec:conclude}
This paper proposes an Inline-enabled Delta-Compression-based approach (DeltaFS) for aggressive write traffic reduction in log-structured file system on mobile devices.
DeltaFS leverages the benefits of large inline area in delta maintenance, and manages the inline area carefully for utmost compressing efficacy.
In addition, DeltaFS introduces a novel main data area delta-maintenance scheme, to break through the constraint of limited inline area.
Finally, DeltaFS advocates the advantages of log-structured file system to mitigate the write amplification issue of delta compression.
Evaluation results show that DeltaFS could reach a significant write traffic reduction by effective data compression.

\section{Acknowledgements}

%This document is modified from the ASPLOS'21 and ASPLOS/22 submission guides, thank
%you Emery Berger, Christos Kozyrakis, Shan Lu and Thomas Wenisch!


\bibliographystyle{IEEEtran}
\bibliography{references.bib}

\clearpage
%\appendix
%\section{First appendix section}
%Perhaps some proofs, extra results, methodologies.

\end{document}